# Wikiometrics: A Wikipedia Based Ranking System

Gilad Katz* and Lior Rokach†

*University of California, Berkeley  † Ben-Gurion University of the Negev
giladk@berkeley.edu  liorrk@bgu.ac.il

**Abstract**

We present a new concept—Wikiometrics—the derivation of metrics and indicators from Wikipedia. Wikipedia provides an accurate representation of the real world due to its size, structure, editing policy and popularity. We demonstrate an innovative "mining" methodology, where different elements of Wikipedia – content, structure, editorial actions and reader reviews – are used to rank items in a manner which is by no means inferior to rankings produced by experts or other methods. We test our proposed method by applying it to two real-world ranking problems: top world universities and academic journals. Our proposed ranking methods were compared to leading and widely accepted benchmarks, and were found to be extremely correlative but with the advantage of the data being publically available.

## 1. Introduction

Ranking is the process by which the relative standing of items is determined. This process is common in multiple domains, both scientific and not. Ranking is considered a difficult problem in many cases as there is no absolute "ground truth" to which the generated ratings can be compared. Nonetheless, multiple studies have been performed that utilize ranking in general and Wikipedia in particular.

Wikipedia has been used in multiple scientific fields: computer science, medicine, physics, sociology etc. According to [1], a growing number of Wikipedia-related papers seems to be generated with each passing year. Wikipedia has several traits which constitute it as such a valuable source of information for research:

- *Size and scope* - As mentioned above, the English Wikipedia alone has over 4.6 million entries. Encyclopedia Britannica, one of the best-known "regular" encyclopedias, has 40,000. This great difference in scope suggests that Wikipedia covers a multitude of fields and areas of interest that are not covered by curated encyclopedias.
- *Timely and updated* – Because of Wikipedia's open editing policy which enables any person to modify its content, the information it contains is almost always up-to-date. Case in point: In 2013, a few minutes after the election of the new pope, one of the authors of this study reviewed the relevant Wikipedia entries and found them to already be updated with the elected pope's new status.
- *Tags and meta-data* – Wikipedia contains multiple types of user-generated content (UGC); categories, links, redirect pages and infoboxes can all be used to infer the type, attributes and connections among the various entities represented in Wikipedia.
- *Wisdom of the crowd* – Since every person has the ability to contribute content to Wikipedia, it reflects the thoughts, ideas and perceptions of peoples, groups and societies [2]. This enables us to use Wikipedia to measure popularity, importance and influence. In a sense, Wikipedia is "representative of the real world."

We argue that Wikipedia's scope and open editing policy render it a representation of the real world. By representation, we mean that the "footprint" of an entity or a concept in Wikipedia is often indicative of its popularity or importance in the real world. It is our belief that by applying this approach to Wikipedia, researchers will be able to use it to address multiple real-world challenges. This change of focus could be significant, as Wikipedia's currently most utilized feature is its text.

In this study we propose a novel concept – *Wikiometrics* – the derivation of metrics and indicators from Wikipedia. While entities ranking is often subjective, we argue that Wikipedia represents the "wisdom of the crowd" and can effectively reflect common perceptions. We propose using three Wikipedia features – infobox data, links and page views – and applying them to the ranking of two of the most widely studied tasks in scientometrics: the ranking of world universities and academic journals. In both cases we compare our results to those obtained by leading and widely-accepted rankings and show that the correlation between our proposed ranking and each of the baselines is similar to the correlation of the baselines among themselves.

Our contribution in this study is twofold: first, we propose a novel approach to a previously unaddressed problem – the ranking of real-world objects. Secondly, we demonstrate how two underutilized Wikipedia features – the *infoboxes* and the *page views* – can be effectively used to address this challenge.

The remainder of this paper is organized as follows. In Section 2 we review related work while in Sections 3 and 4 we present two case studies and evaluate the performance of the proposed methods. In Section 5 we present our conclusions and future research directions.

## 2. Related Work

In this section we review four topics. In Section 2.1 we describe existing ranking methods of world universities. In Section 2.2 we elaborate on the DBpedia project which aims to extract structured content from Wikipedia. In Section 2.3 we go over existing methods for the ranking of scientific journals. Finally, in Section 2.4 we describe the Academic Journals WikiProject, which aims to improve Wikipedia's coverage of scientific publications.

### 2.1. International rankings of world universities

Nowadays, several methods exist which are generally accepted for ranking world universities. In this section we review three such rankings, to which we later compare our proposed ranking methods: Academic Rating of World Universities[1] (ARWU), Times Higher Education World University Ratings[2] (THE) and the Webometrics Rating.[3]

**Academic Rating of World Universities (ARWU)**

The ARWU was the first attempt at establishing a worldwide university evaluation metric. Founded in 2003 by Shanghai Jiao Tong University, its initial goal was to provide a benchmark

---

for Chinese academic institutions. Over the years it has grown in popularity and today it is widely regarded as an accurate measurement tool.

The weights that make up the ranking are as follows (as they appeared on the ranking's official website in May 2013):

1) Alumni of an institution winning Nobel Prizes and Fields Medals – 10%
2) Staff of an institution winning Nobel Prizes and Fields Medals – 20%
3) Highly cited researchers in 21 broad subject categories – 20%
4) Papers published in the journals Nature and Science – 20%
5) Papers indexed in Science Citation Index-expanded and Social Science Citation Index – 20%
6) Per capita academic performance of an institution – 10%

The main advantage of this indicator is its clarity – the ranking method is simple, objective and transparent. On the other hand, its critics claim that it puts too great an emphasis on the natural sciences at the expense of the humanities and that it does not take quality of teaching into account.

**The Times Higher Education World University Ratings**

This rating system is a joint operation conducted by the Times Higher Education magazine[4] and Thomson-Reuters.[5] Its evaluation metrics consist of 13 sub-categories that are grouped into five categories (as of May 2013):

1) Teaching: the learning environment – 30%
2) Research: volume, income and reputation – 30%
3) Citations: research influence – 30%
4) Industry income: innovation – 2.5%
5) International outlook: staff, students and research – 7.5%

This rating is also well accepted and currently considered to be one of the top-three most influential ratings of world universities. It has been criticized for assigning "unfair advantage" to institutions with a small number of undergraduate students and much like the ARWU, it is also criticized for favoring science-oriented universities.

**Webometrics Ranking of World Universities**

This ranking, founded in 2004, attempts to assess the quality of universities based on the volume, visibility and impact of their online content. The ranking is compiled by the Cybermetrics Lab of the Spanish National Research Council and is released twice a year.

The rationale of this ranking system is that a university's quality is reflected in its online presence (i.e., the amount of content available on its domain) and the influence of its publications. This ranking is unique in that it ranks not only several hundreds of well-known universities, but thousands of institutions around the world. The ranking is composed of two categories of ratings, each consisting of 50% of the overall grade:

1) Visibility – The visibility score is calculated by analyzing all links that point to the online content of the evaluated academic institution. This score is defined by the ranking's official website as "…recognizing the institutional prestige, the academic performance, the value of the information, and the usefulness of the services as introduced in the webpages according to the criteria of millions of web editors from all over the world." The creators of the ranking rely on two companies, Majestic SEO[6] and Ahrefs,[7] in order to obtain the relevant link information.

---

[4] http://www.timeshighereducation.co.uk/
[5] http://thomsonreuters.com/

2) Activity – This category is divided into three sub-categories, all with equal weight in the final rating (16.6% of the final rank). These three categories are:
   - **Presence** – The total number of webpages hosted in the university's domain as indexed by Google.
   - **Openness** – The number of recent publications (currently from 2008) that are hosted on the university's web domain, appear on Google Scholar and are publically accessible.
   - **Excellence** – Counts the number of papers that members of the university have published and which are in the *top 10%* of the most cited papers in their respective fields. The data is provided by the Scimago Group.[8]

An extensive analysis conducted by Aguillo *et al.* [3] shows that the greatest dissimilarity among the ratings systems is between the Webometric rating and that of the Times Higher Education. This is not surprising since the former relies heavily on bibliometrics while the latter takes many additional factors into account.

It is important to note that one of our proposed metrics – the infoboxes-based one – is a combination of the ratings presented above. Very much like the Webometrics data, we use data available online in order to rate academic institutions; however Wikipedia is more accessible and more compact than the entire web. For example, it is possible to download all of Wikipedia without the need to use a web crawler.  Like the ARWU and THE, we take into account notable persons (through their Wikipedia pages) who are connected to the university. It should be noted, however, that it is irrelevant to our ranking whether these individuals pursue an academic career or not. Naturally, the differences in emphasis lead to differences in the ratings of various universities and we elaborate on this subject further in the evaluation section.

### 2.2. DBpedia

DBpedia is a collaborative effort to extract structured content from Wikipedia. It is arguably one of the easiest ways to access and utilize structured information extracted from the online encyclopedia. The extracted content is then made available through a database that enables complex queries (e.g., "Which cities in the United States have a population of over 4 million people?").

The project was initiated in 2007 by teams at the Free University of Berlin and the University of Leipzig along with OpenLink Software.[9] DBpedia was released under a free license, enabling others to reuse the code. It currently classifies close to 4 million objects, of which around 2.5 million are part of a consistent ontology. In addition, it contains hundreds of thousands of links to external web pages, close to 200,000 links to other RDF databases (Auer et al., 2007) and over 8 million YAGO categories [4]. The generated ontology is diverse and includes persons, places, creations (music albums, movies, etc.), organizations and many other diverse entities ranging from athletes, to anatomy to television shows. The 359 classes form a shallow ontology that has 1,775 properties.

---

[6] http://www.majesticseo.com/

[7] https://www.ahrefs.com

[8] http://www.scimagoir.com/

[9] *www.openlinksw.com*

Wikipedia's infoboxes are one of the main sources of valuable information. These boxes, located on the right side of many pages (see example in Figure 1) contain structured information that can be easily extracted and added to the database. For example, it is possible to easily discover the predecessor/successor of a particular monarch, length of reign, number of children and numerous other details.

In this study we utilize the information found in the infoboxes to identify universities, faculty members and alumni in an attempt to produce an automatic ranking of world universities. As shown in many cases in the past (and in our evaluation in Sections 3 and 4), the "wisdom- of-the-crowd" can sometimes serve as an excellent substitute for the opinion of experts.

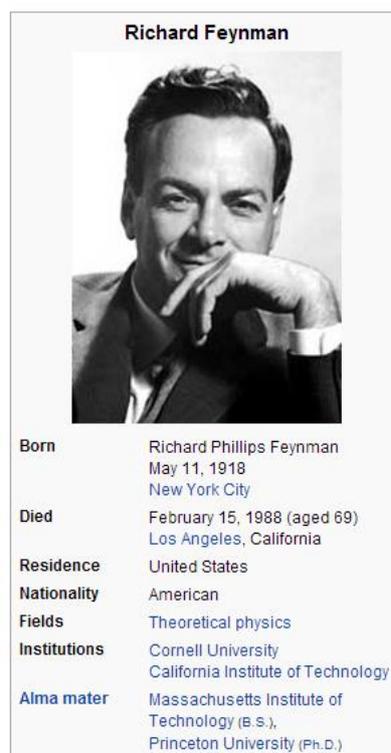

**Figure 1.** An example of a Wikipedia infobox (note the "Alma mater" attribute, which is utilized by one of our proposed ranking methods).

## 2.3. Journal Rankings

Journals serve as the main outlets for publishing the results of scientific research. Journal rankings assist academic libraries in selecting which books and journals to purchase and are often used as a measure of research quality; given a journal's ranking, researchers can target their papers to top-ranked journals and improve their chances for promotion.

The four common approaches to generating journal rankings are *a)* opinion surveys; *b)* citations; *c)* author affiliation and; *d)* behavioral approaches. In expert opinion surveys, a number of scholars rank each journal according to a predefined set of criteria. The results reflect the cumulative peer opinion of a representative group of experts within a particular discipline or field. However, expert surveys have also been criticized for their subjectivity, the lack of clarity of their rating criteria [5], and various biases (such as preferring outlets that publish more articles per year [6]). Finally, establishing a valid expert survey that includes a sufficiently large number of qualitative responders can be time-consuming.

Many citation-based measures have been suggested for ranking journals, including impact factors [7], the Eigenfactor [8], and the h-index and its variants [9]. The main advantage of these measures is their objectivity. However, they have also been criticized, with some claiming that a few highly cited papers can skew the citation distribution [10] or that not all citations have the same significance [5]. Moreover, because citation patterns vary across disciplines, it is very difficult to evaluate multidisciplinary journals. Research shows that using citation-based measures tends to generate journal rankings that are only weakly correlated with expert surveys (see, for instance, [11] and [6] for a complete list). Even when a strong correlation can be found, there are still considerable differences in the ranking of certain journals [12].

The underlying premise of the university affiliation approach is that tenured faculty members of prominent research universities tend to publish their work in premier journals. The Author Affiliation Index (AAI) of a journal (or set of journals) is defined as the percentage of authors who publish in that journal (or set of journals) and are affiliated with a predetermined group of top-rated universities or university departments in the domain under study [13-15].

Behavior-based approaches examine the actual publishing behaviors of tenured researchers at an independently determined set of prominent research universities. This approach assumes that these particular faculty members tend to publish their works in outlets which they regard as being of high quality in the field under study. The behavior of these researchers can be trusted because they have demonstrated a level of research excellence which is recognized by their peers (who have participated in their tenure and promotion committees). Rokach [16] has shown that the publication power approach (PPA) that was developed by [5] for identifying the premier journals can reliably rank AI journals.

### 2.4. The Academic Journals WikiProject

A WikiProject is a general term for a collaborative project undertaken by members of the Wikipedia community. Currently, there are over 2,250 such projects[10] underway, with a large diversity of goals.

The Academic Journals WikiProject is an attempt to improve Wikipedia's coverage of scientific publications by expanding, categorizing, and cleaning up existing articles, as well as creating new ones. This is done by scanning four types of "citation tags" in Wikipedia - {{Citation}}, {{Cite journal}}, {{Vancite journal}}, and {{Vcite journal}}. This task is repeatedly performed by bots.

Although the ranking is informative (and—as we prove later in this study—useful and indicative) it is by no means free from mistakes. The WikiProject's website[11] describes several deficits of its rating system (we present only a few examples here; please see the website for the full list):

- Citations that are not in the formats specified above are not counted.
- Multiple citations on similar pages by the same author are given equal weight to all others.

---

[10] http://en.wikipedia.org/wiki/Wikipedia:Database_reports/WikiProjects_by_changes

[11] http://en.wikipedia.org/wiki/Wikipedia:WikiProject_Academic_Journals/Journals_cited_by_Wikipedia

- Multiple citations of the same journal on the same page will all be counted if the citation format is not identical.

Despite all the above-mentioned deficiencies, our experiments show that the extracted data is still very valuable, although some simple heuristics had to be applied.

## 3. First Case Study: University Ranking

In this section we demonstrate how information extracted from DBpedia and Wikipedia can be used to accurately rank world universities. The proposed method consists of two phases – information extraction and ranking. During the information extraction phase, we query DBpedia and Wikipedia in order to extract a set of entities (academic institutions and persons affiliated with them). We then use this information during the ranking phase in order to rank the extracted university entities.

In order to extract all the relevant entities from Wikipedia, all that was needed was a simple DBpedia query for all entities of type "University." As this query resulted in tens of thousands of institutions we filtered this list, keeping only entities that appeared in *at least* two of the three rankings presented in Section 2.1. This step was also made necessary by the fact that there is large variance in the number of universities in each ranking: ARWU ranks 500, THE ranks only 400 and the Webometrics ranking rates over 12,000 universities. Following this decision, we were left with the task of ranking 389 universities – a large enough number to accurately calculate correlation. In Table 4 we present the top 20 universities ranked by each of the three baseline methods and the most successful variant of our Wikiometrics approaches.

### 3.1. The Proposed Approaches

We propose three distinct methods for the ranking of real world entities: a) *links* – the number of distinct Wikipedia pages which contain links pointing to the entity's Wikipedia page; b) *Overall page views* – the number of times a specific page was viewed over a certain period of time and; c) *Relevant infobox attributes* – the identification of various entities associated with each of the ranked items and the evaluation of their importance. Next, we describe these methods in detail.

### 3.1.1. Links

Let $W$ be the corpus of all Wikipedia entities and $W_r$ be a set of Wikipedia entities (i.e. pages) we wish to rank. For each ranked entity $e$, we count the number of entities in $W$ that contain links pointing to $e$ (see Equation 1). It should be noted that we count the number of *entities*, not links. Therefore, even if one entity contains multiple links to the ranked entity it will only be counted as a single link. This was done in order to prevent a small number of highly-detailed entities from affecting the ranking process.

$$Rank_{links}(e) = \sum_{w \epsilon W} \begin{cases} 1 & w \text{ contains a link to } e \\ 0 & otherwise \end{cases} \quad (1)$$

This ranking method attempts to quantify the importance of an entity by measuring the number of other entities in Wikipedia that choose to cite it. This ranking is based on two hypotheses: a) Wikipedia contributors are more likely to refer to entities whose reputation or importance they judge to be the greatest; b) when choosing which entity to refer to, the first entities that are likely to come to the contributor's mind are those in which he or she holds in highest regard. In essence, this ranking method can be considered as the "Wikipedia version" of the visibility component of the Webometrics ranking method (discussed in Section 3).

We present two variations of this ranking method – the one presented above, which we call *incoming links*, and another we call *incoming-outgoing ratio*. This measure is calculated by dividing the number of incoming links (shown above) by the number of links to other entities in the ranked entity's page (outgoing links). As in the incoming links, multiple outgoing links to the same entity are counted as one. This was done to correct for cases where an entity has associations or connections to multiple entities (collaborations, etc.) that may increase the number of links pointing to it.

**3.1.2. Page Views**

Let $W$ be the corpus of all Wikipedia entities and let $W_r$ be a set of Wikipedia entities we wish to rank. For each ranked entity $e$, we count the number of times it has been viewed throughout a certain period of time. It is important to note that the views of the *redirect pages* (pages which immediately transfer the user to another page) pointing to the entity are also counted in the entity's overcall count.

The hypothesis behind this ranking method is that the more important/prestigious an entity is perceived to be, the more page views it is likely to have. In order to negate the effects of temporary "spikes" in popularity (due to news events, for example) the page views were aggregated over a period of several months.

**3.1.3. Infobox Attributes**

We use the information from the infoboxes in order to obtain for each academic institution the *notable persons* that are associated with it. As Wikipedia's editing policy requires that a person be "notable" to have an entry, we deemed this definition valid for all entities of type "Person" in DBpedia. In addition, we also extracted a general "visibility indicator" for each academic institution. The extracted features are as follows:

- **Faculty members** – For each university, we count the number of people who have at least one of these attributes in their infoboxes: *workInstitution*, *employer* and *workplaces*.
- **Alumni** – We count the notable alumni of each institution by counting the persons with at least one of the following attributes: *alumnus*, *alumna*, *alma mater*, *education* and *training*.
- **Other affiliations** – This component is used to detect additional affiliations that can contribute to the reputation of an academic institution. Here, we counted all persons with at least one of the following attributes: *visitorSchool*, *publisher*, *coachTeams* and *college*
- **Visibility** – The goal of this ranking component is to estimate the "prominence" of each university. It counts the number of articles in which each university's name was mentioned. It should be noted that this attribute is different from the one calculated in Section 2.1 as it does not take into account only links but any appearance of the term. This leads to difference in performance, as is shown later in this section.

Following empiric experimentation and analysis, we chose to use the following formula:

$$Score = 0.5 * WorkInPlace + 0.3 * AlmaMater + 0.1 * TotalSearch + 0.1 * AllRelation$$

This formula resembles some of the components of the leading rating schemes presented in Section 2. As with ARWU, our approach allocates a sizeable part of the ranking to past and current staff. Like the Webometrics rating, we also take into account citations of the university (although not in a purely academic context) and like THE, we incorporate the university's image and connections to the outside world into the rating (although we do not limit these connections to those with industry, as is the case with THE).

**Table 1. The components of the three leading university rankings – Academic Rating of World Universities (ARWU), Times Higher Education (THE) and Webometrics. The weight assigned to each component is in bold.**

| ARWU | THE | Webometrics |
|---|---|---|
| Alumni who are Nobel laureates – **10%** | Teaching: the learning environment – **30%** | Visibility: all links that point to the online content of the institution – **50%** |
| Faculty members who are Nobel laureates – **20%** | Research: volume, income and reputation – **30%** | Presence: the number of webpages in the university's domain – **16.6%** |
| Highly cited researchers – **20%** | Citations: research influence – **30%** | Openness: the number of recent publications that are hosted on the university's domain and appear in Google Scholar – **16.6%** |
| Papers published in Nature & Science – **20%** | Industry income: innovation – **2.5%** | |
| Papers indexed in Science Citation Index-Expanded – **20%** | International outlook: staff, students and research – **7.5%** | Excellence: the number of papers published by faculty which are at the top 10% of their field – **16.6%** |
| Per capita academic performance of the institution – **10%** | | |

**3.2. Results**

We evaluate our proposed ranking method by calculating its correlation to the leading and well-known university rankings reviewed above – ARWU, THE and Webometrics, all from 2011. The Wikipedia version that was used was from December 2013, and the page views statistics were extracted from September-December 2013. Since each ranking method has a different number of universities, we left only universities that appeared in at least two of the three abovementioned rankings (390 in total).

In Table 2 we present the Kendall tau rank correlation between the three existing ranking methods. It is clear that they are highly correlative (all correlations are statistically significant with p<0.001). In Table 3 we present the correlation of our proposed ranking approaches to each of the three original ranking methods. We also show the correlation of each of the components of the proposed methods, as well as that of a *combined ranking* consisting of all three approaches together.

When the ranking method consisted of several components (namely, the Links and Infobox attributes and the Combined option presented in Table 3), we treated the problem of assigning weights to each component as a *rank-based nonparametric regression task* [17]. The purpose of this approach is to maximize the Kendall tau correlation with the dependent variable. As the three original rankings are highly correlative, we only present the results obtained with ARWU as the dependent variable (the results with the other rankings as the dependent variable are highly similar). The Nelder-Mead method [18] is used to assign the optimal values.

Table 2. The Kendall tau correlation of the three "original" university ranking methods – Academic Rating of World Universities (ARWU), Times Higher Education (THE) and Webometrics – with each other. All the correlations are statistically significant (p<0.001).

|   | ARWU | THE | Webometrics |
|---|---|---|---|
| **ARWU** |  | 0.565 | 0.467 |
| **THE** |  |  | 0.435 |
| **Webometrics** |  |  |  |

Table 3. The Kendal tau correlation of our proposed rankings (in grey) and each of their components (in white) to the three original universities rankings. All correlations were found to be statistically significant (p<0.001).

|   | ARWU | THE | Webometrics |
|---|---|---|---|
| **Links** | 0.372 | 0.396 | 0.475 |
| Incoming Links | 0.375 | 0.474 | 0.396 |
| Incoming-outgoing links ratio | 0.295 | 0.387 | 0.295 |
| **Page views** | 0.357 | 0.435 | 0.423 |
| **Infobox Attributes** | 0.498 | 0.451 | **0.485** |
| Alma Mater | 0.402 | 0.45 | 0.427 |
| Faculty Members | **0.525** | 0.468 | 0.475 |
| Other Affiliations | 0.389 | **0.477** | 0.406 |
| Visibility | 0.388 | 0.387 | 0.42 |
| **Combined** | 0.501 | 0.468 | 0.477 |

All the correlation results presented both in Tables 2 and 3 are statistically significant (p<0.01). These results are a clear indication that the three approaches presented in this paper can be used as simple and effective tools for ranking. In addition, it is clear that the more complicated method – the infobox attributes ranker – shows the best performance. This is not surprising, as its features were "handcrafted" for this ranking task.

An interesting observation from Table 3 is that the *Faculty Members* component alone enables us to reach a correlation that is close or even better than the one obtained by combining *all* prediction methods together. This is an important observation that attests to the simple truth that the researchers of an institution (past and present) are the most important element in determining its quality.

Finally, we would like to emphasize an important point: The statistical test used to determine ranking similarity assigns equal weight to all items on the list. This means that our proposed approaches fared well not only for the top 20 or top 100 world universities (which may be considered easier to rank) but also for lower-ranking universities.

**Table 4: The top 20 universities according to Wikiometrics and the three "benchmark" methods.**

| Rank | Wikiometric | ARWU | Webometrics | Times Higher Education |
|---|---|---|---|---|
| 1 | Harvard University | Harvard University | Harvard University | California Institute of Technology |
| 2 | Massachusetts Institute of Technology | Stanford University | Massachusetts Institute of Technology | Harvard University |
| 3 | University of California, Berkeley | Massachusetts Institute of Technology | Stanford University | Stanford University |
| 4 | University of Cambridge | University of California, Berkeley | University of California, Berkeley | University of Oxford |
| 5 | University of Oxford | University of Cambridge | Cornell University | Princeton University |
| 6 | Princeton University | California Institute of Technology | University of Michigan | University of Cambridge |
| 7 | Stanford University | Princeton University | University of Minnesota | Massachusetts Institute of Technology |
| 8 | Yale University | Columbia University | University of Washington | Imperial College London |
| 9 | Columbia University | University of Chicago | University of Wisconsin – Madison | University of Chicago |
| 10 | University of Chicago | University of Oxford | University of Texas at Austin | University of California, Berkeley |
| 11 | University of Michigan | Yale University | University of Pennsylvania | Yale University |
| 12 | Cornell University | University of California, Los Angeles | Pennsylvania State University | Columbia University |
| 13 | University of Pennsylvania | Cornell University | Columbia University | University of California, Los Angeles |
| 14 | University of Toronto | University of Pennsylvania | Carnegie Mellon University | The Johns Hopkins University |
| 15 | The Johns Hopkins University | University of California San Diego | University of Illinois at Urbana–Champaign | Swiss Federal Institute of Technology in Zurich |
| 16 | California Institute of Technology | University of Washington | University of California, Los Angeles | University of Pennsylvania |
| 17 | University of California, Los Angeles | University of California San Francisco | Texas A&M University | University College London |
| 18 | University of Wisconsin – Madison | The Johns Hopkins University | University of Maryland College Park | University of Michigan |
| 19 | New York University | University of Wisconsin – Madison | Purdue University | University of Toronto |
| 20 | University of Illinois at Urbana–Champaign | University College London | University of North Carolina at Chapel Hill | Cornell University |

### 3.3. Analysis and Discussion

As it is clear that the infobox-based approach fared best out of the three approaches, we chose to perform an in-depth analysis of its performance. We begin our evaluation of the results by analyzing the ranking distribution. Then, in addition to the overall comparison presented in the previous section, we perform an additional analysis using only North American universities. The reason for this is the (relative) increased visibility of these institutions in the English Wikipedia—a fact which may skew the results. It should be noted again that all results reflect the ratings that were published in 2011.

We began by analyzing the score distribution produced by our proposed approach. The results are presented in Figure 2. The purpose of this analysis was to determine whether the scores could be fitted to a known parametric distribution. In particular, the null-hypothesis that the new measure is distributed as log-normal is accepted using the chi-square test with p-value=0.27062.

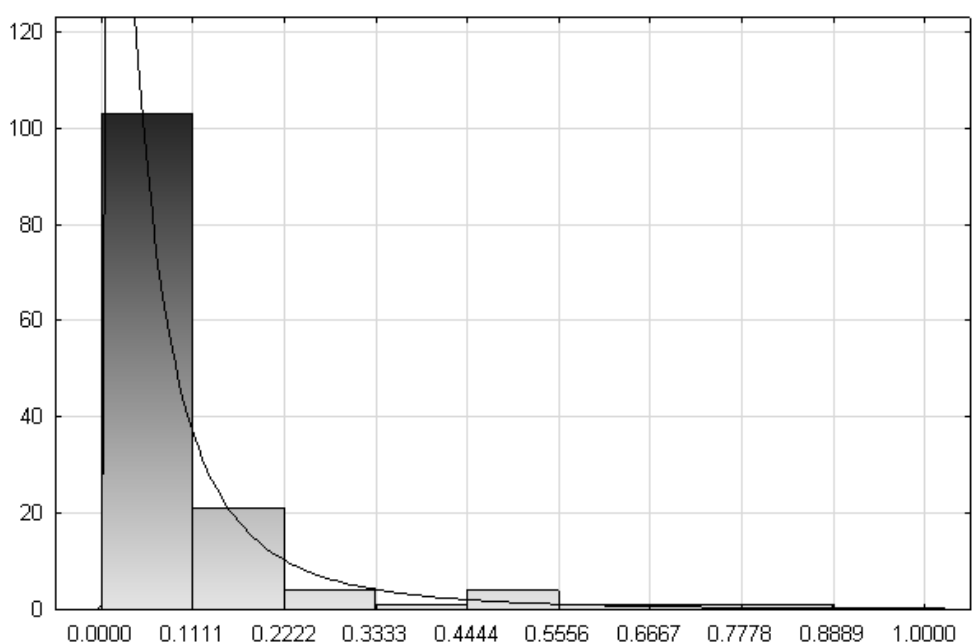

**Figure 2:** The distribution of the proposed measure and the fit to log-normal.

Next, we present a scatter diagram of the rankings of the Wikiometrics and the Times Higher Education rankings. Figure 3 illustrates the log-fit correlation of the two measures, which was found to be statistically significant with r=0.6550. The visualization clearly demonstrates that the two rankings are very similar throughout the entire range of scores. Similar results were obtained with the other two baseline ranking methods.

**Evaluating North American Universities**

Since North America is home to many of the world's top universities, it could be argued that rankings such as Webometric and the one proposed in this paper are affected by the fact that these universities have a much larger Wikipedia (and overall web) exposure than those of other countries. This possible bias is presented in Table 5, which shows the number of North American universities in the top 100 and top 200 of the analyzed rankings. It can easily be seen that both the Webometrics and Wikiometrics methods are those with the largest number of North American universities.

It should be noted that this bias exists despite the fact that DBpedia, which is used in this research, includes localized versions of Wikipedia in 111 languages. This might be the case due to the fact that the English version of Wikipedia is far richer than other language versions. For example, the English version contains 763,643 entries for persons while the French version contains only 62,942. Naturally, each language has a better coverage of prominent researchers expressing themselves in that particular language. By combining these two aspects (richness of the English version and the preference for those speaking one's own native language), it is to be expected that native English speakers will be relatively better covered.

For all the reasons mentioned above, we decided to also calculate the correlation of the various rankings while including **only** North American universities. The results are presented in Table 6 and clearly show that the correlation of Wikiometrics with existing rankings becomes even higher. This may indicate that Wikiometrics is more reliable when implemented in each area separately. While it might be worthwhile to mitigate the "over-exposure" of North American universities (by normalizing the counts, for example), we leave this issue for future research.

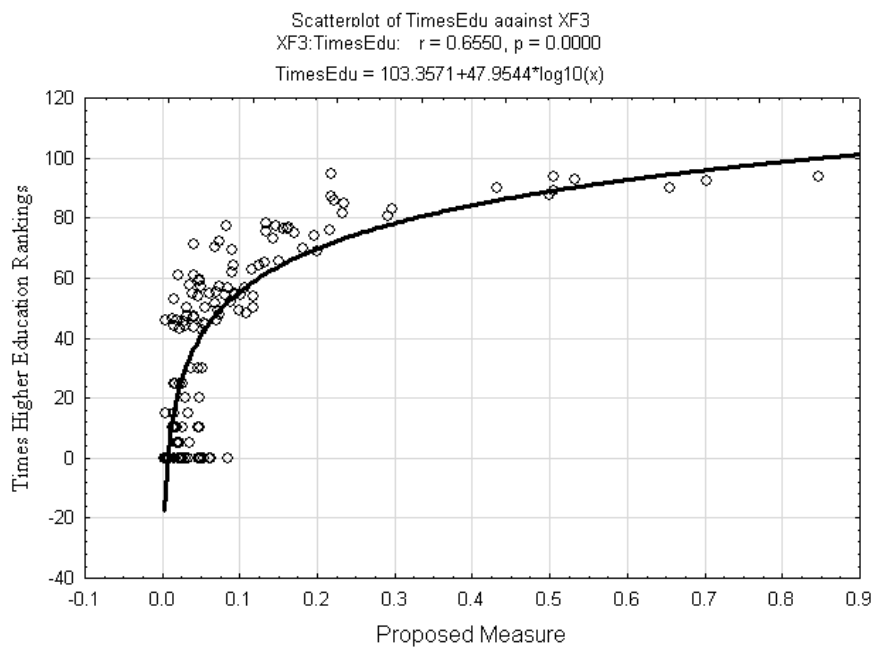

**Figure 3:** Comparing the values of the proposed measure and the Times Higher Education measure.

**Table 5: North-American Universities in the Top 100 and 200 of our proposed method and the three "benchmark" methods**.

| Ranking | Top 100 | Top 200 |
|---|---|---|
| Times Higher Education | 56 | 85 |
| Shanghai | 57 | 87 |
| Webometrics | 74 | 109 |
| Wikiometrics | 60 | 98 |

Table 6: The Spearman correlation results of the proposed method (and each of its components) to well-known ranking methods for North American universities. All scores indicate correlation with p<0.01.

|  | ARWU | THE | Webometrics | Wikiometrics |
|---|---|---|---|---|
| **ARWU** |  | 0.680223 | 0.680341 | 0.859634 |
| **THE** | 0.680223 |  | 0.747693 | 0.801552 |
| **Webometrics** | 0.680341 | 0.747693 |  | 0.773092 |
| **Wikiometrics** | 0.859634 | 0.801552 | 0.773092 |  |

## 4. Second Case Study: Journal Ranking

In this section we demonstrate that Wikipedia can be used to great effect for ranking academic journals. This task has proven more difficult than the ranking of top world universities, as many journals are not represented in the "regular" Wikipedia, but only in projects such as the WikiGroups described in Section 2.4. This makes two of the approaches presented in the previous case study – the use of links and page views – not applicable. Therefore, we address this task by utilizing a weighted set of infobox attributes.

We chose to focus on the artificial intelligence (AI) domain in order to compare our outcome to previously published results. Several rankings of AI journals are available in the literature. Cheng et al. (1996) and Serenko (2010) used citation-based measures while Serenko and Dohan (2011) reported on expert surveys in the field. Rokach (2012) used author-based rankings.

### 4.1. Information Extraction

As mentioned in Section 2.4, we utilized information from the Academic Journals WikiProject group as the basis of our ranking method. The measures used in this case study were:

1. Citations: Number of times the journal was cited by Wikipedia
2. Citers: Number of Wikipedia articles that have cited the journal (if the same Wikipedia article cited the journal twice then it was counted only once)
3. Has Wikipedia Page – This is a binary indicator which gets the value of 1 if the journal has a dedicated Wikipedia page. Our assumption is that top-tier journals have a dedicated Wikipedia page (this, however, does not apply for all journals and is the reason for not using the links and page views ranking methods).

In addition, we created a linear combination of all the above measures. In order to find the weights we used linear regression with the dependent (target) variable being the 5-year impact factor that was published by the Thomson-Reuters.

As mentioned in Section 2.4, the values generated by Wikiometrics have two serious limitations. The first is that it cannot account for citations that are not annotated by the recommended "<REF>" tag. The second limitation is that even the slightest variation in name will be considered as a separate journal. An example of this problem is the journal *IEEE Transactions on Patterns Analysis and Machine Intelligence*. This journal appears in 13 different variants, including: *PAMI, PATTERN ANALYSIS AND MACHINE INTELLIGENCE*, and *IEEE Transaction On*. Consequently, we used simple heuristics in order to aggregate entries that refer to the same journal.

### 4.2. The Ranking Phase

Our evaluation encompassed 108 peer-reviewed AI journals that were identified according to the sub-category "Computer Sciences – Artificial Intelligence" as indexed by the Thomson-Reuters Web of Knowledge (WoK). This data refers to all journal publications of the benchmark scholar.

We used linear regression in order to find the optimal weights for the abovementioned parameters. After scaling the values of Citation and Citers to [0,1], we arrived at the final formula:

$$Score_{journal} = 4.3848 * Citers + 4.42 * Citation + 0.8238 * HasPage \quad (3)$$

The measure that was used to evaluate the correlation of our proposed method to existing rankings was the Spearman rank order correlations. The results of the evaluation are presented in Table 7 (Correlations marked with an asterisk are significant at $p < 0.05$). As in the previous section, all results are based on rankings published in 2011.

Analysis of the results clearly indicates the existence of a high correlation between Wikiometrics and existing rankings. Moreover, Wikiometrics was the only ranking method to be correlative with all other rankings with a $p<0.01$. While being correlative with all existing rankings (Power; 2-year Impact Factor; 5-year Impact Factor; Expert Survey Score) the highest correlation was found with the Expert Survey Score.

**Table 7: The Spearman correlation values of our proposed method and other well-known journal ranking methods (statistically significant results are marked with an asterisk (*)).**

|  | Wikiometric Citers | Wikiometric Citations | Wikiometric HasPage | Wikiometric Combined | Power | 2 years Impact Factor | 5 years Impact Factor | Expert Survey Score |
|---|---|---|---|---|---|---|---|---|
| Wikiometric Citers |  | 0.991451* | 0.407459* | 0.935816* | 0.533244* | 0.486636* | 0.534429* | 0.713065* |
| Wikiometric Citations | 0.991451* |  | 0.404296* | 0.932995* | 0.504644* | 0.488743* | 0.535447* | 0.667878* |
| Wikiometric HasPage | 0.407459* | 0.404296* |  | 0.660699* | 0.199476 | 0.221053 | 0.253653* | 0.324381* |
| Wikiometric Combined | 0.935816* | 0.932995* | 0.660699* |  | 0.485119* | 0.477281* | 0.529575* | 0.666104* |
| Power | 0.533244* | 0.504644* | 0.199476 | 0.485119* |  | 0.221165 | 0.223009 | 0.517432* |
| 2 years Impact Factor | 0.486636* | 0.488743* | 0.221053 | 0.477281* | 0.221165 |  | 0.941599* | 0.533394* |
| 5 years Impact Factor | 0.534429* | 0.535447* | 0.253653* | 0.529575* | 0.223009 | 0.941599* |  | 0.581586* |
| Expert Survey Score | 0.713065* | 0.667878* | 0.324381* | 0.666104* | 0.517432* | 0.533394* | 0.581586* |  |

### 5. Conclusions and Future Work

In this paper we presented two case studies that demonstrate Wikipedia's ability to provide valuable information regarding the real world, by capitalizing on the "wisdom-of-the-crowd" and extracting simple metrics from it. We showed that the opinions of tens of thousands of people (if not more) could constitute a surprisingly accurate alternative for the opinions of experts.

We believe that the estimates provided by Wikiometrics could be further improved by a more elaborate processing of its contents. For example, for the journal ranking one can also take into account references that do not use the <Ref> template and extend the evidence used for the ranking. To this end, an appropriate references extraction would have to be developed for the correct identification of the journal title. The university rankings could be improved in a similar way by taking into account the university affiliation of notable individuals even if this information is not indexed in the infobox but appears as biographical text.

We are currently considering several directions for future work. The first direction is the attempt to generate Wikiometrics ratings for additional domains, particularly those completely unrelated to academia. We hypothesize that Wikiometrics might serve as a low-cost means for reliably measuring issues of interest. For example, Wikipedia could be used for ranking movies and even to predict box office success [19].

An additional direction is the use of other types of attributes – network centrality and graph analysis – for the challenges presented in this paper. Finally, we also consider "automating" the Wikiometrics ratings by implementing a machine learning approach that would utilize different facets and attributes of Wikiepedia in order to produce rankings on any domain specified by the user.